\documentclass{sig-alternate-2013}

\newfont{\mycrnotice}{ptmr8t at 7pt}
\newfont{\myconfname}{ptmri8t at 7pt}
\let\crnotice\mycrnotice%
\let\confname\myconfname%

\usepackage{flushend}
\usepackage{color}
\usepackage{url}
\usepackage[super]{nth}
\usepackage{multirow}
\usepackage{rotating}
\usepackage[flushleft]{threeparttable}
\usepackage{amssymb}

\permission{Permission to make digital or hard copies of all or part of this work for personal or classroom use is granted without fee provided that copies are not made or distributed for profit or commercial advantage and that copies bear this notice and the full citation on the first page. Copyrights for components of this work owned by others than ACM must be honored. Abstracting with credit is permitted. To copy otherwise, or republish, to post on servers or to redistribute to lists, requires prior specific permission and/or a fee. Request permissions from permissions@acm.org.}
\conferenceinfo{HT'14,}{September 1--4, 2014, Santiago, Chile.}
\copyrightetc{Copyright 2014 ACM \the\acmcopyr}
\crdata{978-1-4503-2954-5/14/09\ ...\$15.00.\\
http://dx.doi.org/10.1145/2631775.2631808}

\begin{document}

\title{Online Popularity and Topical Interests\\ through the Lens of Instagram}
 
\numberofauthors{2}

\author{
Emilio Ferrara\\
\affaddr{School of Informatics and Computing}\\
\affaddr{Indiana University Bloomington, USA}\\ 
\email{ferrarae@indiana.edu}
\alignauthor
Roberto Interdonato, Andrea Tagarelli\\
 \affaddr{DIMES}\\
 \affaddr{University of Calabria, Italy}\\ 
\email{\{rinterdonato,tagarelli\}@dimes.unical.it}
 }

\maketitle

\begin{abstract}
Online socio-technical systems can be studied as proxy of the real world to investigate human behavior and social interactions at scale. Here we focus on Instagram, a media-sharing online platform whose popularity has been rising up to gathering hundred millions users. Instagram exhibits a mixture of features including social structure, social tagging and media sharing. The network of social interactions among users models various dynamics including follower/followee relations and users' communication by means of posts/comments. Users can upload and tag media such as photos and pictures, and they can ``like'' and comment each piece of information on the platform. 
In this work we investigate three major aspects on our Instagram dataset: \emph{(i)} the structural characteristics of its network of heterogeneous interactions, to unveil the emergence of self organization and topically-induced community structure; \emph{(ii)} the dynamics of content production and consumption, to understand how global trends and popular users emerge; \emph{(iii)} the behavior of users labeling media with tags, to determine how they devote their attention and to explore the variety of their topical interests. 
Our analysis provides clues to understand human behavior dynamics on socio-technical systems, specifically users and content popularity, the mechanisms of users' interactions in online environments and how collective trends emerge  from individuals' topical interests.
\end{abstract}

\section{Introduction}
The study of society through the lens of social media allows us to uncover questions about human behavior at scale \cite{lazer2009life}. Recent results unveiled complex dynamics in human behavior \cite{vespignani2009predicting,centola2010spread}, interactions \cite{aral2009distinguishing,crandall2010inferring} and influence \cite{aral2012identifying,bond201261}. 
Still, many open questions remain: for example, how do social interactions affect individual and collective behavior? Or, how does connectivity affect individual and collective topical interests? Yet, how do trends and popular content emerge from individuals' interactions? 

In this paper we address these questions by studying an emerging socio-technical system, namely Instagram. The popularity of this platform has been growing during recent years: as of the beginning of 2014 Instagram gathers over one hundred million users. Instagram users generate an unparalleled amount of media content. Hence, it should not be surprising that Instagram has recently attracted the attention of the research community, fostering results in different areas including cultural analytics \cite{hochman2012visualizing,hochman2013zooming} and urban social behavior \cite{silva2013comparison}.
Instagram represents an unprecedented environment of study, in that it mixes features of various social media and online social networks (including the ability of creating user-generated content in the form of visual media), the option of social tagging, and the possibility of establishing social relations (\emph{e.g.},  followee/follower relationships), and social interactions (\emph{e.g.},  commenting or liking media of other users.) 

A natural comparison arises between Instagram and other photo sharing systems, particularly Flickr. 
The two systems appear rather different in terms of features and target of users. Flickr offers more professional-oriented features (\textit{e.g.}, high-quality photos, thematic groups and communities, advanced media organization features.) Instagram, being designed for mobile users, resembles an amateur photo-blog, as it incorporates features to quickly take photos and apply visual effects, and it offers a minimal interface. In other words, Flickr can be seen as a more complete photo sharing platform with social network features, while 
Instagram resembles a Twitter-like online social network based on photo sharing. 

Following the lead of studies based on similar platforms such as Flickr \cite{rattenbury2007towards,crandall2009mapping,mislove08wosn,cha2009www}, in this paper we address five different research questions, discussed in the following, spanning different areas of network-, semantic- and topical-based data analysis using signals from user activities and interactions. 

\subsection{Contribution and outline}
We provide a framework to analyze the Instagram ecosystem, incorporating in our model the unique mixture of social interactions, social tagging and media sharing features provided by the platform. 
By using this framework, we conduct a rigorous analysis focusing on the following main aspects: 
\textit{(i)} the structural characteristics of the Instagram network, 
\textit{(ii)} the dynamics of content production and consumption, and 
\textit{(iii)} the users' interests modeled via the social tagging mechanisms available to label media with topical tags. 
We elaborate on each and all these aspects to answer the following research questions: 
 
\begin{description} 
\item[\textbf{Q1}] \textit{Network and community structure}:\ 
What are the salient structural features in the network built on the users' interactions? 
\item[\textbf{Q2}] \textit{Content production and consumption}:\ 
How do users produce and consume content? That is, how do users get engaged on the platform and how do they interact with content produced by others? 
\item[\textbf{Q3}] \textit{Social tagging}: 
How diverse is the set of tags exploited by each user? In other words, what is the user tagging behavior?
\item[\textbf{Q4}] \textit{Topical clusters of interest}: \ 
How can users be grouped based on the tags they use to annotate media?
\item[\textbf{Q5}] \textit{Popularity and topicality}: \ 
How does the topical interests of users affect their popularity? And, how large is the variety of topics covered by each user or by each media?
\end{description} 

\subsection{Scope of this work}
To the best of our knowledge, this work is the first to study the Instagram network of users' interactions, social tagging activities, and topical interests. Therefore, our major goal is to fill a lack of knowledge concerning a number of research issues in Instagram. Within this view, we aimed at providing a first understanding of the above listed aspects of the Instagram network, being aware that all such aspects are interrelated and hence they should be preferably addressed together. It should also be noted that our experimental findings depend on the particular sampling mechanism used to build our  dataset; as we shall discuss in the next section, this introduces a bias that does not allow us to provide an  analysis of the full Instagram ecosystem, but only of users (and associated media) that are engaged in a public Instagram initiative. 

\section{Methodology}

In this section we describe the challenges that we faced in gathering data from the Instagram network, and the technical choices that we adopted to build our dataset. 
Analogously to other studies, we had to cope with the impossibility of obtaining data directly from the network administrators; therefore, we collected an Instagram sample by querying the Instagram API.\footnote{See \url{http://instagram.com/developer/}} 
Various features are made publicly available, including: \emph{(i)} the \emph{users API}, which allows sampling from the Instagram user space by querying for specific user account details; \emph{(ii)} the \emph{relationships API}, which retrieves information about specific users, their followers and followees; \emph{(iii)} the \emph{media API}, which queries for specific or popular media; \emph{(iv,v)} the \emph{comments} and the \emph{likes} APIs, respectively, to extract comments and likes from specific media; and \emph{(vi)} the \emph{tags API}, which extracts the keywords associated with specific media, as attributed by the social tagging process of Instagram users.



\begin{table}\centering
\caption{Statistics on the Instagram media dataset.}
\begin{tabular}{|l|l|}
\hline\hline
No. Media		&	 1,686,349 \\
No. Distinct users		&	 2,081 \\
No. Tags			&	 8,919,630 \\ 
No. Distinct tags& 	269,359 \\
No. Likes		&	 1,242,923,022 \\
No. Comments 	& 	 41,341,783 \\ 
\hline\hline
\end{tabular}
\label{tab:statistics}
\end{table}

\subsection{Crawling strategy}
Our primary objective in crawling the Instagram network was to ensure adequate levels of consistency in user relationships as well as topical variety in media properties, over a timespan possibly larger than the actual crawling period. We expected to detect a user interaction graph having topological properties (\textit{e.g.}, clustering coefficient, average path length) as close as possible to those typically exhibited by other (directed) social media networks \cite{WilsonBSPZ09,MisloveMGDB07}; at the same time, we aimed at collecting media whose thematic subjects could span over a predetermined, relatively large classification, while capturing time information about media and user relationships that would allow for trend evolution analysis. 
 
Our initial crawling attempt consisted in retrieving media geolocalized w.r.t. a list of touristic/popular locations, which were selected based on their presumed potential to attract users with very different (photographic) tastes, concerning, \textit{e.g.}, art and culture, entertainment and night life, wild life (sea/mountain), etc.  Then, the user relations underlying the authors of the retrieved media were   taken into account to build a user network. Our hypothesis here was that two users who take pictures within a limited area are more likely to be connected via a follower/followee relation (they may know each other in real life.) 
Unfortunately, despite the spatial proximity between the authors of the collected media, a poor number of followships were identified, resulting in a network overly disconnected (\textit{e.g.}, clustering coefficient of 2.0E-6). Note that, by trying different sets of touristic locations, we obtained similar results in terms of connectivity.

We changed our crawling strategy based on retrieving users that belong to a relatively large ``community'' in Instagram. Here, our usage of term community corresponds to that of thematic channel, which is typical in many other social media networks (\textit{e.g.}, YouTube); Instagram 
does not offer an explicit group/community 
feature, therefore we exploited the existence of public initiatives officially organized by Instagram. We focused our crawling on the Instagram \emph{weekend hashtag project} (WHP) promoted by the Instagram's official blog.\footnote{\url{http://blog.instagram.com/tagged/weekend-hashtag-project}} 
The characteristics of the WHP initiative and their implications on our data crawling are described next.

 \subsection{Dataset construction}
Every Friday, the Instagram team runs a photographic contest, through the Instagram's official blog. Each contest is assigned a specific topic, which is expressed by a unique (hash)tag prefixed with \emph{\#whp}. 
According to the project rules, submitted photos need to be marked with no more than one contest-specific tag. 

We selected $72$ popular contests and randomly picked up about $2,100$  users that participated in at least one of those contests. All media uploaded by these users (including media that were not tagged with \emph{\#whp}-hashtags) were gathered and their information retrieved and stored into the \textit{media dataset}.
For each media, we retrieved its unique ID, the ID of the user who posted it, the timestamp of media creation, the set of tags assigned to the media, the number of likes and comments it received. 

We constructed the \emph{Relational Instagram Network} (RIN) as a directed weighted graph. Edges were drawn to model asymmetric relationships of the form follower-followee, and edge weights were calculated proportionally to the number of likes and comments generated by a user (follower) towards media created by her/his followee. 
The users selected to build the media dataset were used as seed nodes for the construction of the RIN. 
Note that we conceived the RIN so to model (asymmetric) relationships that hold strictly among the participants in the contests. The reason for this choice is that including the whole topological neighborhood of the candidate nodes (\textit{e.g.}, the individual egonets also including non-participants) would have resulted again in highly disconnected networks (with clustering coefficient of the order of 1.0E-6). 
Therefore, we started a breadth-first search process from the set of seed nodes, filtering out any user  who did not participate in at least a \emph{\#whp} contest.
 %
 
Our data were crawled over about one-month period (from Jan 20 to Feb 17, 2014). 
The obtained media dataset contains full information about over 2 thousand users and  almost 1.7 million media, with about 9 million tags, 1.2 billion likes, and  41 million comments (see Table~\ref{tab:statistics}.) 
Details on our RIN are reported in Table~\ref{tab:network_statistics}. 
Here it can be noted that the network of user relations shows a negative, close-to-zero assortativity, which would indicate no tendency of users with similar degree to connect each other. Moreover, the characteristic path length and clustering coefficient are both low, while the modularity is rather high, which would indicate that the RIN has small modules, with moderately dense connections between the nodes within modules and sparse connections between nodes in different modules.\footnote{Our data are available at \url{http://uweb.dimes.unical.it/tagarelli/data/}.}

\begin{table}\centering 
\caption{Relational Instagram network statistics.}
\begin{tabular}{|l|l|}
\hline\hline
No. Nodes 				& 44,766 \\
No. Links 				& 677,686 \\ 
Avg. In-degree 		& 15.14 \\
Avg. Path length 		& 3.16 \\ 				
Clustering coefficient 	& 4.1E-2 \\
Diameter 				& 11 \\
Assortativity index 	& -0.097 \\ 
No. Communities			&	151 \\
Network modularity		&	0.578 \\
\hline\hline
\end{tabular}
\label{tab:network_statistics}
\end{table}

\paragraph{Limitations} 
As previously discussed, our dataset is intentionally built around the set of users and media that belong to a competition-driven, large, community in Instagram. 
Unlike previous work on the Flickr network (a major competitor of Instagram)~\cite{mislove08wosn}, we were not able to perform a number of analyses such as, \textit{e.g.}, preferential creation/reception and proximity bias in link creation, which rely on fellowship creation timestamps. This information is missing in our dataset, as the Instagram APIs do not make it available. Flickr APIs do not make it available either, but those authors inferred such temporal information by crawling the Flickr network daily, and monitoring the creation of new links~\cite{mislove08wosn}.  
Another limitation concerns the analysis of latent interactions (\textit{e.g.}, profile browsing), which has been shown to be a prominent activity in OSNs~\cite{BenevenutoRCA09,SchneiderFKW09,JiangWWHSDZ10}: unfortunately, this information is not publicly available for Instagram, while obtaining significant clickstream data (like that used other studies~\cite{BenevenutoRCA09,SchneiderFKW09}) is challenging. 

\section{Analysis and Results}


\begin{figure}[!t]
\includegraphics[width=\columnwidth]{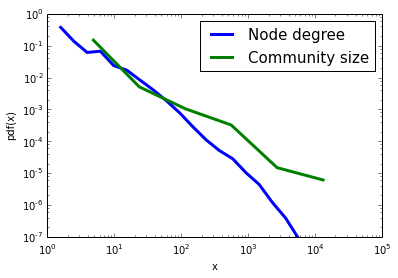}
\caption{Distribution of node degree and community size of the Relational Instagram Network.}
\label{fig:node_cs_distribution}
\end{figure}

We begin with explaining the five research questions that we will address to unveil the characteristics of Instagram. 

\subsubsection{Q1: Network and community structure}
Our first  question aims at understanding what are the structural features of the Relational Instagram Network and the characteristic of its community structure. 
We want to determine the dynamics of social relations and interactions on the system and how they shape (if they do) the structure of the network. In addition, we want to determine whether or not the community structure reflects the self-organization principle \cite{luhmann1995social} by which individuals in social networks tend to aggregate in communities oriented to topical discussions, and if this, in turn, yields the emergence of a 
\emph{topically-induced community structure}.

\subsubsection{Q2: Content production and consumption}
We want also to understand how the cycle of production and consumption of information (\emph{e.g.}, media) is characterized on Instagram. We first aim at understanding what is the driving mechanism of content production; then, we aim to unveil whether content consumption, measured in some way (\emph{e.g.}, via social interactions), follows similar patterns or if any striking difference emerges.

\subsubsection{Q3: Social tagging dynamics}
In the third research question our goal is to study the dynamics of social tagging on Instagram. We want to study both the patterns of tag adoption at the user level, and at the global level, to characterize how popular tags emerge from the adoption of independent users.
We also want to describe the variety of tagging usage by the users, to determine whether users focus their attention on few rather than many contexts.

\subsubsection{Q4: Topical clusters of interests}
A fourth research questions aims at determining whether it is possible to cluster users exploiting   their tagging behavior, and, in turn, if topical clusters emerge by means of such procedure. 

\subsubsection{Q5: Popularity and topicality}
Our final research questions aims at unveiling the dynamics of user popularity and how this relates to topical interests. We hypothesize that popular users might exhibit different patterns of attention and therefore different topical interests. We want to determine whether we can characterize user popularity as function of the variety of their interests, and, in turn, learn how topicality relates to social interactions. 

\begin{figure}[!t]
\includegraphics[width=\columnwidth]{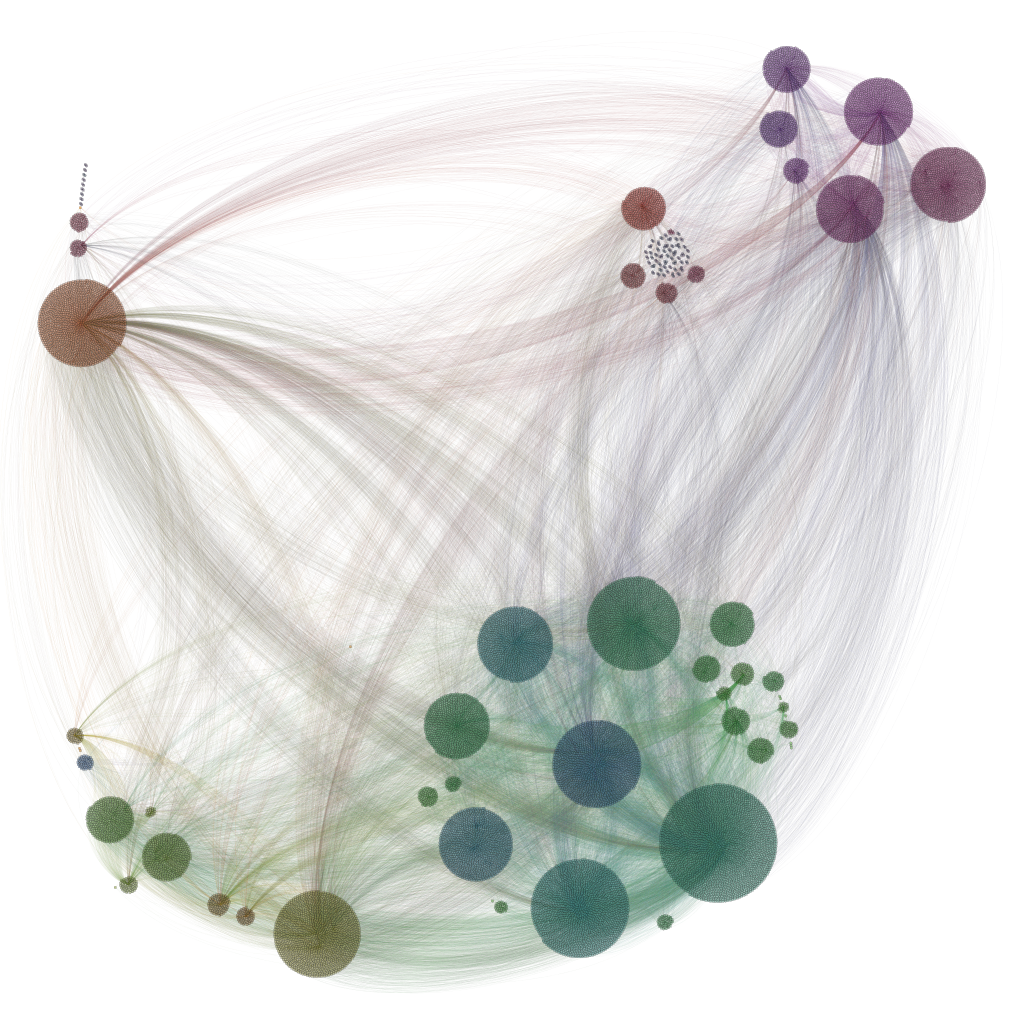}
\caption{Visualization of the community structure of the Relational Instagram Network.}
\label{fig:cvis_community}
\end{figure}

\subsection{Structural features of the Instagram Network}

\begin{figure}[!t]
\includegraphics[width=\columnwidth]{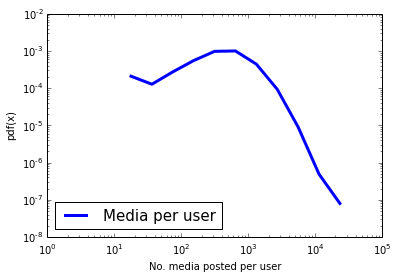}
\caption{Distribution of user content production.}
\label{fig:media_distribution}
\end{figure}

We discuss the analysis of the Relational Instagram Network (RIN) we carried out to answer our first research question (\textbf{Q1}).  
Our goal here is to study its topological features and determine whether they reflect any particular social process.
In particular, we aim at unveiling whether this particular environment, at the boundary between a social network and a sharing media platform, exhibits any characteristic feature: for example, we will drive our attention on the effect of topical interests of users and how these reflect on the network structure. 
Figure \ref{fig:node_cs_distribution} shows the distribution of node degree (in blue) and community size (in green) for the RIN. 
The community detection task has been carried out using two algorithms: the \emph{Louvain method} \cite{blondel2008fast}, and OSLOM \cite{lancichinetti2011finding}. Results obtained with both methods are consistent (the plot shows the results from the former algorithm.) 
Both the node degree and the community size distributions are broad and exhibit a fat-tail.
A broad degree distribution suggests that the Relational Instagram Network growth may follow a preferential-attachment mechanism \cite{barabasi1999emergence}: new social relations and social interactions are disproportionately more likely to occur between individuals who previously grew their social network and invested in interacting with others, rather than between users less prone to connect \cite{simon1955class}. 
The formation of communities of heterogeneous size suggests the emergence of self organization \cite{luhmann1995social}, a principle explaining that individuals tend to aggregate in units (the communities) optimized for efficiency of communication (\emph{e.g.}, around specific topics of conversation.) 
A self-organized network structure enjoys crucial properties, including that of enhancing the topicality of interests, or their scope, to smaller sets of individuals rather than to the entire system.
By addressing research questions \textbf{Q2} and \textbf{Q3} in the following sections, we will determine whether these communities emerge from user relations and interactions around certain topics of interest; in other words, we will investigate whether the network exhibits a \emph{topically-induced community structure}.

To visualize the community structure of the RIN we produced a graphical representation in Figure \ref{fig:cvis_community}, by means of a circular hierarchical algorithm.\footnote{Cvis by Andrea Lancichinetti: \url{https://sites.google.com/site/andrealancichinetti/cvis}.}
Here nodes (\emph{i.e.}, users) belonging to the same community have the same colors, and the hue of the edges transitions from the color of the community of the source node to that of the target one.
The RIN community structure clearly separates close clusters of individuals (\emph{e.g.}, bottom-right ones) from clusters of isolated individuals (\emph{e.g.}, top-right ones.) 
Note that the RIN has (multi)edges weighted by means of social relations and interactions (\emph{i.e.}, follower/followee, likes and comments), being these weights accounted in the community detection and visualization tasks.
Differently from other social networks \cite{ferrara2012large,grabowicz2012social}, Instagram does not exhibit a tight core-periphery structure, whereas communities of large size exist in peripheral areas of the network and they are interconnected with other communities of comparable size.
Other basic statistics of the Relational Instagram Network are reported in Table \ref{tab:network_statistics}.

\begin{figure}[!t]
\includegraphics[width=\columnwidth]{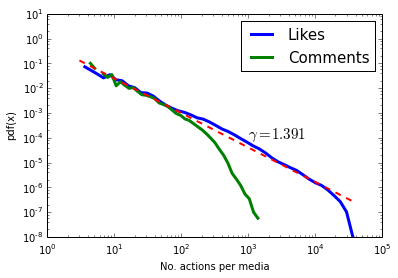}
\caption{Distribution of social interactions.}
\label{fig:actions_distribution}
\end{figure}

\subsection{Content production and consumption}
We continue our analysis of the Instagram ecosystem by investigating how users produce and consume content (\textbf{Q2}.) Our goal is to determine whether any particular pattern emerges to describe how individuals' get engaged on the platform and how they interact with content produced by others.
To this aim, we study \emph{content production} from the user perspective. Figure \ref{fig:media_distribution} shows the probability density function (pdf($x$)) of the amount $x$ of media posted by each Instagram user in our dataset.
This plot suggests peculiar content production dynamics on Instagram: users who already uploaded a large number of media are more likely to do so, causing the presence of a fat tail showing users with a disproportionate amount of media posted on the platform.
Individuals exhibit higher tendency to posting new content if they already did that in the past.
The lack of a scale-invariant content production dynamics differentiates Instagram from other platforms \cite{mislove08wosn} (even if some caution is required given how the sample was constructed.) If our observation holds in general, this has an interesting impact from the perspective of system design, in that it suggests a neat separation between active and inactive users: those who are already engaged in using the platform are more likely to keep staying active users. Strategies to engage inactive users could be designed and implemented based on these findings to lower the heterogeneity (\emph{i.e.}, the imbalance) in users involved in content production on the platform.

We now investigate \emph{content consumption} on Instagram. Here with content consumption we intend that a given user on the platform has performed some specific action toward a media produced by another user (\emph{e.g.},  liking or commenting it.) This draws an interesting parallel between content production and social interactions, and provides a slightly different perspective from usual studies on platform like social media such as Twitter, where content consumption is intended as users rebroadcast others' content (\emph{e.g.}, via retweets) aiming at information diffusion rather than interactions.
Figure \ref{fig:actions_distribution} shows the distribution of two consumption dynamics, namely ``like'' and comment, of Instagram users.
The plot includes the best fit of a power law to the likes distribution,\footnote{The statistical significance of this fitting (and all the others in the paper) has been assessed by means of \emph{powerlaw}, a library by Alstott \emph{et al.} \cite{alstott2013powerlaw}, and it's based on a Kolmogorov-Smirnov test.}  with an exponent $\gamma=1.391$ ($x_{min} = 3$, $\sigma = 0.001$), whereas no significant power law fit has been found for the comment distribution that clearly shows two different regimes, $x\lessapprox 250$ and $x\gtrapprox 250$.
The ``likes'' distribution shows a cut-off in the tail due to the finite system size, and suggests that the behavior of likes and comments on Instagram might follow two different dynamics. 
Popularity of media measured by the number of likes grows by preferential attachment similarly to how, for example, scientific papers acquire citations \cite{jeong2003measuring}: resources with large number of  likes (resp., citations) are more likely to acquire even more.
Differently, the ecosystem is less prone to trigger large conversations (based on comments); this is consistent with the theory of user communication efficiency: the different costs (\emph{e.g.}, in terms of time required to perform the action) between ``liking'' some content and writing a comment affect the nature of interactions among individuals on the platform.

\begin{figure}[!t]
\includegraphics[width=\columnwidth]{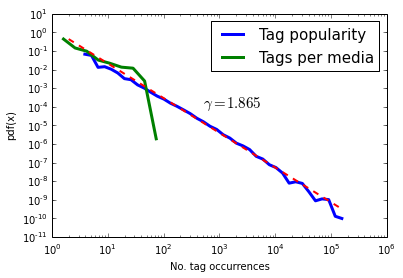}
\caption{Tag adoption and global popularity.}
\label{fig:tags_distribution}
\end{figure}

\subsection{Social tagging dynamics} \label{sec:social_tagging}

To answer our question about the dynamics of social tagging on Instagram (\textbf{Q3}) we investigated three aspects: \emph{(i)} the tag popularity at the global level and the distribution of tags per media; \emph{(ii)} the distribution of total tags used by the users and their vocabulary size; and, \emph{(iii)} the diversity in tag usage by each individual.

Our first goal is to understand how tags emerge in the system at the global level from the tagging patterns of individual media.
To this end, we derived the distribution of tag popularity, as represented by the probability density function of observing a given total number of tag occurrences across all media.
Then, we obtained the distribution of the number of tags assigned to each media. 
The results are shown in Figure \ref{fig:tags_distribution}. The plot reports the best fitting of a power law to the distribution of tag popularity with an exponent equal to $\gamma=1.865$ ($x_{min} = 2$, $\sigma = 0.002$), whereas the tags-per-media distribution best fits an exponential-decay function.  
Two main observations stand out. First,  the tag usage mechanism seems to follow an information economy principle of least effort, that is that the majority of media are labeled with just a few tags, and larger sets of tags assigned to the same media are increasingly more unlikely to be observed.
Second, although the mechanism describing the assignment of tags is not quite by preferential attachment, the outcome of the process, that is the overall tag popularity, follows a power law behavior. 
Similar findings have been observed in other popular systems, like Twitter, where popular (hash)tags emerge from individuals' adoption \cite{weng2012competition}. Limited attention of users and competition among (hash)tags have been hypothesized as explanation of the emergence of such broad distributions.

\begin{figure}[!t]
\includegraphics[width=\columnwidth]{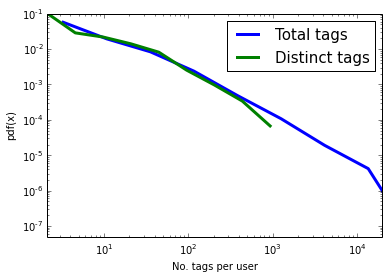}
\caption{Tag usage and tagset size distributions.}
\label{fig:user_tags_distribution}
\end{figure}

Moreover, we seek to understand what is the emerging behavior at the user level.
We want to determine what patterns of tag adoption users follow, in terms of how many total tags they use, and how many of these tags are distinct. In other words, we establish their vocabulary size (\emph{i.e.}, the number of ``words'' they are aware of) and we compare it against the total number of tags they produce.
Figure \ref{fig:user_tags_distribution} shows the distribution of, respectively, total and distinct tags used by each user. Both distributions are fat-tailed and show similar slopes. 
Vocabulary size reflects the information economy principle: the distribution of distinct tags per user spans above one order of magnitude less if compared with that of the total tags usage. This suggests that the actual user vocabulary size is limited, with a large majority of users adopting only few tags. This can be explained by considering that users cannot keep track of all tags emerging on the platform.

Finally, to the aim of studying how diverse is the set of tags used by each individual we proceeded as follows.
First, we described each given user $u$ in our dataset by means of a vector $T_u$ where each entry represents the frequency $f(t)$ of adoption of tag $t$ (\emph{i.e.}, the total number of times user $u$ adopted tag $t$ to label one of the media she/he uploaded to Instagram), for all tags used by $u$.
We define the entropy value $H(\cdot)$, to describe each user's entropy in the adoption of tags, in the classic Shannon way

\begin{equation*}\small\label{eq:user_tag_entropy}
H(u) = -\sum_{t\in T_u}{p(t)\cdot \log p(t)}, \quad \mbox{with } p(t)=\frac{f(t)}{\sum_{t\in T_u}{f(t)}}.
\end{equation*}

\begin{figure}[!t]
\includegraphics[width=\columnwidth]{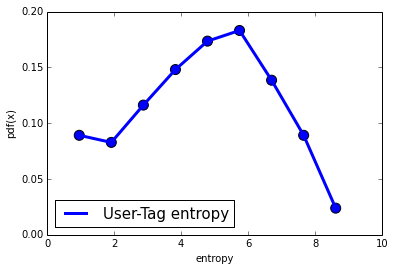}
\caption{User-Tag entropy distribution.}
\label{fig:user_tags_entropy}
\end{figure}

Afterwards, we determined the probability density function of the distribution of users' tag adoption entropy, as shown in Figure \ref{fig:user_tags_entropy}.
Note that the entropy ranges between $0$ and the logarithm of the total number of tags of each user. The lower the entropy, the more focused a user's tagging pattern is (that is, she/he tends to adopt less tags in a more concentrated ways),  the more diverse is her/his tagging behavior.
Figure \ref{fig:user_tags_entropy} shows that the entropy is roughly normally distributed with a peak between $5$ and $6$, and a skewness towards lower values of entropy. 
This suggests that, while a fraction of about 50\% of the users tend to exhibit an average tagging variety (corresponding to entropy values $4\lessapprox x\lessapprox 7$), the remainder are either focused ($x\lessapprox 4$) or extremely heterogeneous ($x\gtrapprox 7$) in their tagging adoption.
The analysis of tag adoption entropy reveals crucial features from the perspective of modeling user attention: tagging entropy is a proxy to measure how spread or focused users' attention is towards few or several contexts.
A more refined analysis, that will take into account not only tags but the topics that emerge from their co-occurrences is presented later to address \textbf{Q5}.

\subsection{Topical clusters of interest}

To answer \textbf{Q4}, we conducted a number of experiments 
aimed at evaluating how users in the media dataset can be grouped together. 
Users were represented as term-frequency vectors in the space of media tags. 
We performed the clustering of these users based on Bisecting $k$-Means~\cite{steinbach2000comparison}, which is well-suited to produce high-quality (hard)
clustering solutions in high-dimensional, large datasets~\cite{zhao2004empirical}.  
We used the CLUTO clustering toolkit\footnote{CLUTO: \url{www.cs.umn.edu/~karypis/cluto}}  which provides a globally-optimized version of Bisecting $k$-Means.  
Feature selection was carried out to retain only the features (\emph{i.e.}, tags) that accounted for 80\% of the overall similarity of clusters.  
 We experimented by varying the number $k$ of clusters from 2 to 50, with unitary increment of $k$ at each run. Our evaluation was both quantitative, based on standard within-cluster and across-cluster similarity criteria, and qualitative, based on the cluster characterization in terms of descriptive and discriminating features.   The best-quality clustering solution corresponded to $k=5$.

\begin{figure}[!t]
\includegraphics[width=0.45\textwidth]{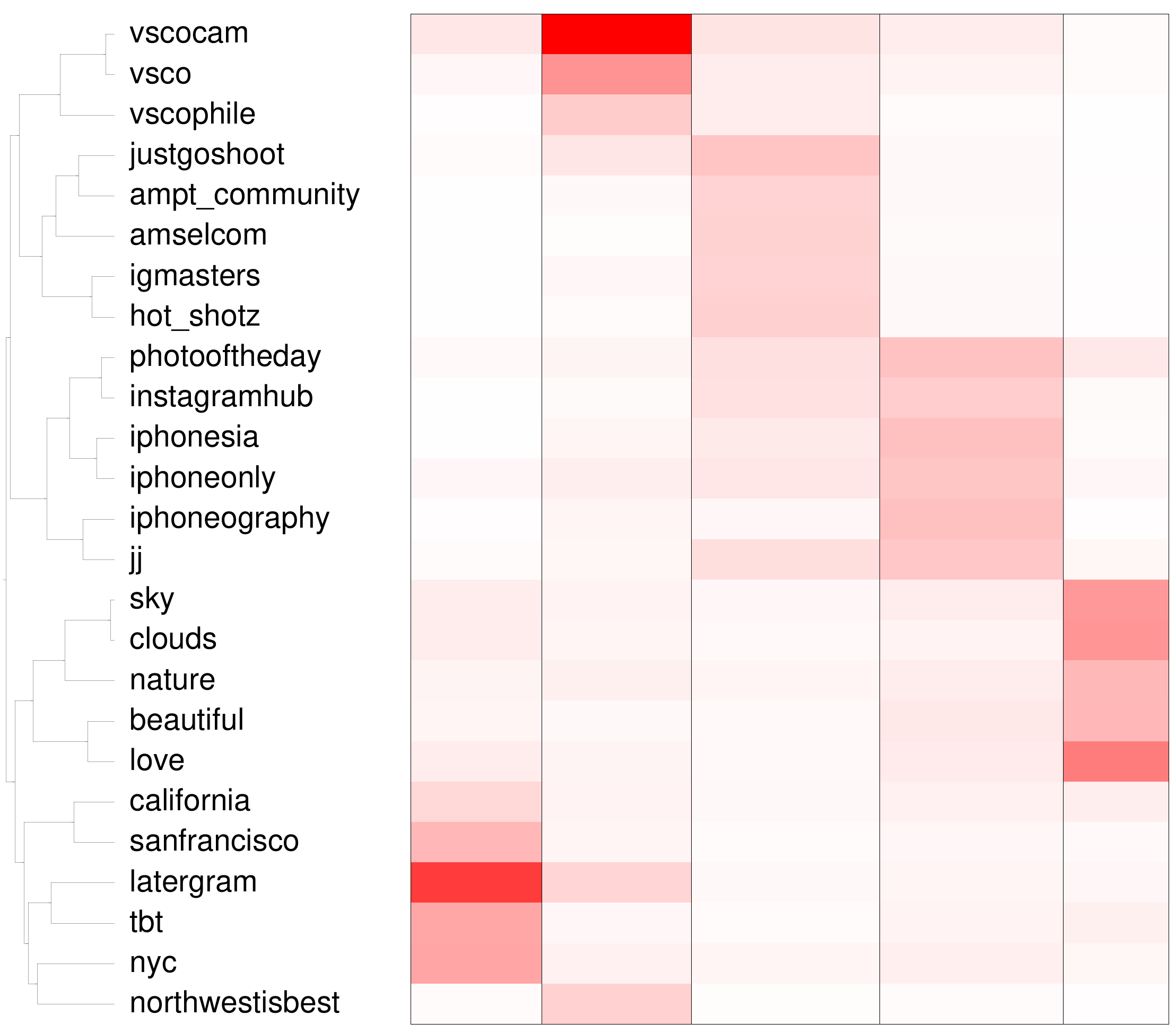}
\caption{5-way clustering of the users in the media dataset.}
\label{fig:topics_5_users}
\end{figure}

Figure~\ref{fig:topics_5_users} shows a color-intensity plot of the relations between the different clusters of users and features (\emph{i.e.}, tags), corresponding to a 5-way clustering solution. Only a subset of the features is displayed, which corresponds to the union of the most descriptive and discriminating features of each cluster. Moreover, features are re-ordered according to a hierarchical clustering solution, which is visualized on the left-hand side of the figure. 
A brighter red cell corresponding to a pair feature-cluster indicates higher power of that feature to be, for that cluster, descriptive (\emph{i.e.}, the fraction of within-cluster similarity that this feature can explain) and discriminating (\emph{i.e.}, the fraction of dissimilarity between the cluster and the rest of the objects this feature can explain.) 
The width of each cluster-column is proportional to the logarithm of the corresponding cluster's size. 

It can be noted that the five clusters are quite well-balanced. The first two clusters (\emph{i.e.}, the two left-most columns) are strongly characterized by hashtags denoting the use of popular \textit{applications}, namely VSCO Cam and Latergram. The former is commonly used to modify pictures by adding filters, while the latter is used to schedule the upload of a picture at different (later) time than that of its shot. The \#latergram cluster is also characterized by another popular hashtag, \#tbt, which is an acronym of  Throwback Thursday (a ``throwback'' theme can pertain to some event that happened in the past), and at higher levels in the induced feature-cluster hierarchy, by geographical hashtags (\textit{e.g.}, \#nyc, \#california.) 
While the fifth cluster is labeled by \textit{subject-based} tags that are evocative of feelings (\#love) or nature (\#sky, \#nature), the third and fourth clusters are instead characterized by either \textit{attention-seeking} tags or \textit{microcommunity-focused} tags: 
\#photooftheday, \#igmaster are representative of the former category, as users are seeking approval from their peers, whereas   \#amselcom, and \#justgoshoot fall into the latter category along with  \#iphonesia (originally used by East-Asia users who share photos taken with their iPhones) and \#instagramhub (which aims at helping users   understand best practices and sharing tips.) Yet, \#jj, which is run by prominent Instagram user Josh Johnson, denotes a community which asks their users to abide by the rule ``for every one photo posted, comment on two others and like three more''. 
Note that, in general,  members of such microcommunities are  often asked to share photos on a specific theme, and motivated to create more effective images. These challenges posed by the community continuously prompt their members to play  active roles in Instagram.


\subsection{User popularity and topicality}
Our final research question (\textbf{Q5}) aims at exploring the topical interests space of users and how this affects their popularity.
To learn the topics of interest exhibited by the users we employed a topic model which adopts the tags assigned by users to their media as the topical characterization feature.
We filtered out tags occurring only once in our corpus, that account for roughly 20\% of the total.

\begin{figure}[!t]
\includegraphics[width=\columnwidth]{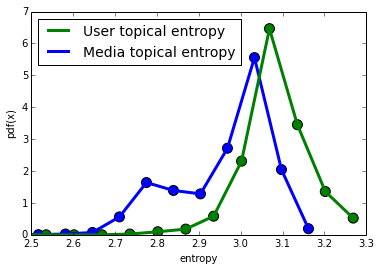}
\caption{User and media topical entropy.}
\label{fig:topic_entropy}
\end{figure}

After experimenting with various topic models available in the \emph{gensim} \textsl{python} library,\footnote{Gensim: \url{http://radimrehurek.com/gensim/}} (including LDA and HDP), we adopted Latent Semantic Indexing (LSI) that provided the most interpretable model for a suitable number of topics set to $10$. 
Note that, differently from topic modeling applications where the impact of the choice of the number of topics might affect the results, in our case we obtained consistent results by using larger number of topics as well (we tested with 5, 10, 20 and 30 topics obtaining consistent results.)
We set up our topic model inferring the posterior probability distribution over the topics for each media in our dataset. To determine the topical interests of each user $u$, we simply averaged the probabilities of each topic being exhibited by the media produced by $u$.
As concerns the variety of topics covered by each user (as well as that exhibited by a given media), we adopted the Shannon entropy. Similarly to the formula used in Section~\ref{eq:user_tag_entropy}  for users and tags, we    calculated the probability of observing the topics (rather than the tags.) 
Afterwards, we estimated the probability distribution of user (respectively, media) topical entropy, as illustrated in Figure \ref{fig:topic_entropy}.
Here we observe that the topical entropy (both for users and media) is very concentrated and spans values between $2.5$ and $3.5$ as opposed to the broader entropy interval of user tags, which ranges between $0$ and $9$ (see Figure \ref{fig:user_tags_entropy}.) This suggests that, although users are equally likely to adopt either a narrow or broad vocabulary of tags, their topical interests tend to be in general more concentrated. At the end of this section we will discuss if there are deviations from this pattern, and how they relate to users' popularity. In other words, we will seek to understand whether popularity can be described by variety  of topical interests.
Note that user topical entropy and media topical entropy are similarly distributed, as it should be, suggesting the goodness of our approach to build users topical interest profiles.

\begin{figure}[!t]
\includegraphics[width=\columnwidth]{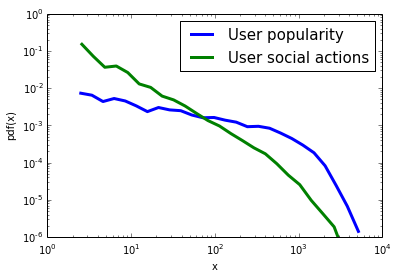}
\caption{User popularity and social actions.}
\label{fig:populary_actions_distribution}
\end{figure}

In order to investigate the popularity of users, we  measured   the total number of likes and comments received by a user's  media. We also account for the total number of times a user likes or comments someone else media, namely the number of \emph{social actions} that this user performs. Such measures are clearly correlated since one is complementary to the other.   
In Figure \ref{fig:populary_actions_distribution} we show the distribution of user popularity and user social actions. 
From the two distributions some interesting facts emerge. First, they are both broadly distributed. 
The slope of the user popularity distribution is small. This implies  the presence of   many users with approximately the same (small) popularity. Around $x\gtrapprox 1000$, the slope of the user popularity distribution drastically changes, becoming steeper as of identifying a cut-off due to the finite size of the sample. Values larger than this point coincide with the few extremely popular users who receive a lot many likes and comments to their media.
The social actions distribution is still broad but with a steeper slope. 
This implies that there exist relatively less users (with respect to the popularity distribution) who produce many likes or comments to others' media.

\begin{figure}[!t]
\includegraphics[width=\columnwidth]{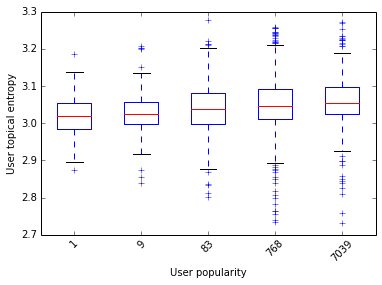}
\caption{Boxplot on popularity and topical entropy.}
\label{fig:gmm_entropy_popularity}
\end{figure}

Our final experiment aims to understand whether user popularity can be explained by means of variety of users' topical interests.
Our goal is to determine whether different classes of popular users emerge, according to their topical interests.
To this aim, we correlate user popularity with their topical entropy values discussed above.   
%
Figure \ref{fig:gmm_entropy_popularity} shows a boxplot that separates users in five logarithmic bins. For each bin, the corresponding box extends from the lower to upper quartile values of the data, whereas the whiskers extend from the box to show the range of the popularity values for that bin. A red line corresponds to the median value for each bin. Popularity once again is measured as the sum of likes and comments received from the media produced by each user. Results do not vary when considering  the count instead of the sum of social actions, or when varying the number of topics in the topic model. The values of topical entropy span between $2.7$ and $3.3$ bits, in a spectrum of $0.6$ bit overall. 

From Figure \ref{fig:gmm_entropy_popularity} two interesting findings emerge. First, user popularity is somewhat affected by topical entropy. As popularity grows, the topical entropy increases accordingly. For example, the median topical entropy for very popular users ($768 < x \leq 7039$) is around $0.1$ bits larger than that of unpopular users ($x \leq 9$).
By comparing these two distributions we observe a statistically very significant difference: a two-sided t-test of the two independent samples yields a t-statistic  of $3.674$ corresponding to a p-value of $0.0005$.
The second observation is that various outliers are present among the popular users; this causes the presence of popular users with topical entropy much lower or much higher than average.

Our findings suggest that unpopular users tend to be more focused in their interests with respect to more popular users. However, there exist popular users who are either extremely specialized (very low values of topical entropy) or have extremely broad topical interests.
These results complement the intriguing hypothesis, recently advanced by other studies \cite{weng2013virality,weng2014predicting}, that popularity might be affected by structural features and information diffusion patterns in addition to content production and topical interests.

\section{Discussion}
In this section we summarize the results obtained addressing the five research questions we posed at the beginning of the paper, providing a final memorandum to the reader with the main findings of this work. 

\begin{description}
\item[\textbf{A1}] \textit{Network and community structure}:\ 
The network structure of the Relational Instagram Network exhibits two relevant characteristics: a scale-free distribution of node degree and a broad distribution of community size. This suggests that the network growth might happen by preferential attachment, whereas the emergence of the community structure might be driven by self organization of users around topics of interest. 

\item[\textbf{A2}] \textit{Content production and consumption}:\ 
Regarding content production,   the life-cycle of information generation on the platform might be explained in Simon's terms with a heightened likelihood that already engaged users produce more content.  
Content consumption, on the other hand, might be driven by the information economy principle of least effort: users tend to adopt the ``like'' behavior strikingly more than producing comments, in line with the intuition that a greater effort (in terms of time and communication economy) must be done to drive the social interaction towards conversation.

\item[\textbf{A3}] \textit{Social tagging}: \
Tag usage too is in line with the principle of least effort: the majority of media are labeled with just a few tags, yielding a power law distribution of tagging activity. A similar effect was recently observed in other platforms, like Twitter~\cite{weng2012competition} due to limited attention in combination with competition among tags.

\item[\textbf{A4}] \textit{Topical clusters of interest}: \ 
Clusters of  Instagram users can be detected by means of the tags they adopt to label the  contents they produce (and how contents are produced), to indicate their intention to seek approval from other users, or to denote the microcommunity the users belong to. 

\item[\textbf{A5}] \textit{Popularity and topicality}: \ 
User popularity is mildly affected by the breath of topical interests. 
Increasing values of topical entropy are positively correlated to higher user popularity; however, popular users exhibit more extreme topical entropy values, which means that some popular users are highly specialized, whereas other have very broad interests. This translates in the principle that users with general interests have the same chance to become popular than the specialized ones.

\end{description}

\section{Related Work}

In recent literature, social media and online communities have been used as proxy to study human communication and behavior at scale in different scenarios, 
including social protests or mobilizations, social influence and political interests and much more~\cite{bond201261,conover2013geospatial,conover2013digital}. 
Other research highlighted how trends emerge and diffuse in socio-technical systems, and how individuals' interacting in such environments devote their attention \cite{leskovec2009meme,budak2011structural}.

In this work, we addressed popularity and trends emerging in Instagram. 
Trends are used to represent popular topics of interest as they are considered indicators of collective attention~\cite{lehmann2012dynamical}, and 
 have been studied to detect exogenous real-world events~\cite{sayyadi2009event,becker2011beyond,ferrara2013traveling}. 
%
Our work explores network features and diffusion patterns of social media content. 
Information diffusion and the network structure in social media have been extensively studied \cite{lerman2010information,grabowicz2012social,weng2013role,de2013analyzing}.
A lot of attention has been devoted to explore the community structure of such socio-technical systems \cite{ferrara2012large} and to study the formation and evolution of social groups therein \cite{backstrom2006group}.
The interplay between the dimensions of social interactions and those of topical interests of users have been also investigated showing a mutual reshaping based on mutual feedback mechanisms \cite{romero2013interplay}. 
Moreover, our study touches on a mixture of ingredients commonly exhibited by socio-technical systems that digitally mediate communications among individuals: content, topics of discussion, language and tags. Content in social media well reflects socio-economic indicators of users, in that languages highlight patterns of linguistic homogeneity \cite{quercia2012social}, 
individual and collective satisfaction, demographic characteristics \cite{mislove2011understanding}. %
Social network data also exhibit cues of users' evolution, as discussed in this work. Existing literature witnesses that online content reflects the intuition that users are susceptible to changing their behavior along with experience, and common patterns of evolution emerge over time \cite{danescu2013no,mcauley2013amateurs}. 

Narrowing our focus on research investigating social media features similar to those of Instagram, 
an extensive analysis of the Flickr social network is reported in~\cite{mislove08wosn,cha2009www}. 
Particular attention here is devoted to the understanding of the temporal evolution of network topology, picture popularity, and relating processes of information propagation. However, unlike in our work, no content information (\textit{e.g.}, tags, comments) is taken into account. 
A study concerning user interactions in the 22 largest regional networks on Facebook is conducted in~\cite{WilsonBSPZ09}. 
Results show that interaction activity on Facebook is significantly skewed towards a small portion of each user's social links, and consequently the interaction graph reveals different characteristics than the corresponding social graph. Note that we also leverage the importance of user interactions, as our RIN takes into account like and comment actions. 
In~\cite{BenevenutoRCA09}, clickstream data obtained by an aggregator of social networks (Orkut, Myspace, Hi5 and LinkedIn) are exploited to analyze various aspects of online lifetime of users, such as frequency and length of online sessions, activity sequences and types, dynamics of social interactions. 
The analysis of workloads has showed that the user browsing is the dominant behavior (accounting for $92\%$ of all requests.) 
A similar study is performed in~\cite{SchneiderFKW09}, where clickstream data from Facebook, LinkedIn, Hi5, and StudiVZ are used to characterize actual user interactions within the sites, in terms of session length, feature popularity and active/inactive time. In our work, we do not consider clickstream data as such information is not made available through the Instagram APIs. 
User latent interactions in Renren are investigated in~\cite{JiangWWHSDZ10}. Results show that latent interactions (\textit{e.g.}, browsing of users profiles) are more numerous, non-reciprocal and they often connect non-friend strangers if compared to visible ones. In contrast to 
previously discussed works, the authors in~\cite{JiangWWHSDZ10} did not use clickstream data, but they exploited public data about visits to the crawled user profiles. 


\section{Conclusions}
In this paper we presented a broad analysis of the Instagram ecosystem, exploiting its heterogeneous structure, part social network, part tagging environment, and part media sharing platform. 
We exploited users signals in the form of relationships and interactions to investigate a number of research questions spanning network-, semantic- and topical-based analysis on users, media and how these two dimensions are interrelated.

We first focused on the network and community structure, observing that the topical interests exhibited by the users might affect their inter-connectivity and interactions, shaping the network structure around topical communities that can possibly be explained by users' self organization.
We then studied the patterns of content production and consumption on the platform, putting into evidence a strong heterogeneity in the mechanism of production of new information, and the emergence of an information economy principle in the case of content consumption.
Our analysis shifted toward the study of social tagging behavior, and we highlighted that users exhibit vocabularies of limited size reflecting their limited attention capabilities, but, nonetheless, popular trends emerge. This can be explained by limited attention of individuals and competition among ``memes'' (\emph{i.e.}, popular tags.)
The study concludes by investigating topics and topicality in the network, and relating it to user popularity. We showed that clusters of   users can be found around the tags.
Furthering this analysis by learning a topic model on such tags, we showed that the variety of topical interests mildly affects user popularity: users with narrow interests tend to be less popular, whereas broader interests tend to yield higher popularity. However, popular users are special in a way because they exhibit more extreme behavior: they can produce either very topically specific content, or media of very broad interest. 

Further work is needed to assess what role the structure of the network has in the determination of the popularity of content and users in online ecosystems based on social connectivity and content sharing, in the direction of recent work on Twitter \cite{weng2013virality,weng2014predicting}.


\end{document}